\documentclass[12pt,a4paper]{article}

\usepackage[tbtags]{amsmath}
\usepackage{amsfonts}
\usepackage[left=1.8cm,right=1.8cm,
    top=2cm,bottom=2cm,bindingoffset=0cm]{geometry}

\usepackage{pgfplots}

\begin{document}

\title{Active beam splitting attack applied to differential phase shift quantum key distribution protocol}
\author{A.\,S.~Avanesov$^1$\footnote{E-mail:daypatu@rambler.ru}, D.\,A.~Kronberg$^2$\footnote{E-mail:dmitry.kronberg@gmail.com}, A.\,N.~Pechen$^{2,3}$\footnote{E-mail:apechen@gmail.com}}
\date{}
\maketitle

$^1$ Moscow Institute of Physics and Technology, Dolgoprudny, Moscow Region, Russia

$^2$ Department of Mathematical Methods for Quantum Technologies, Steklov Mathematical Institute of Russian Academy of Sciences, Gubkina str. 8, Moscow 119991, Russia

$^3$ The National University of Science and Technology "MISiS", Leninsky Prospekt 4, Moscow 119991, Russia

\abstract{The differential phase shift quantum key distribution protocol is of high interest due to its relatively simple practical implementation. This protocol uses trains of coherent pulses and allows the legitimate users to resist individual attacks. In this paper, a new attack on this protocol is proposed which is based on the idea of information extraction from the part of each coherent state and then making decision about blocking the rest part depending on the amount of extracted information.}

\section*{Introduction}

One of the most important problems of cryptography is the secure key distribution, i.e. an exchange of secrete key between two legitimate users (typically called as Alice and Bob). The practical implementation of the secure key distribution would make possible information-theoretic secure communications by using one-time-pad cipher. In 1984 the first quantum key distribution (QKD) protocol BB84 was proposed~\cite{BB84}. Unlike classical cryptography, QKD does not use any assumptions about the computational power of an eavesdropper (called as Eve). Eve can perform any actions, except of those which are not allowed by quantum mechanics like discriminating between two non-orthogonal states (see for example~\cite{Holevo_article,Holevo}). Thus, the security of QKD is guaranteed by the laws of quantum mechanics.

There are several schemes for QKD introduced so far and for some of them the security proofs are known. For instance, full proofs of the security for BB84 protocol were given in~\cite{Renner_security}. QKD protocols with pseudorandom bases were analyzed in~\cite{FedorovPRA2018}. But practical implementation of QKD protocols faces some problems connected with impossibility or difficulty to meet conditions of actual security proven QKD protocols. For example, the detectors used by the receiver are not ideal. Additionally, the real communication channel looses some signals and in combination with the fact that usually the emitted signals are weak coherent pulses, it imposes some restrictions on exploitation of QKD protocols. For instance, the use of weak coherent pulses makes the protocol vulnerable to such new attacks like photon number splitting (PNS) attack~\cite{PNS_attack} or unambiguous state discrimination (USD) attack~\cite{USD}. In these attacks Eve in principle can first obtain full information about some part of the key without causing any disturbance and then block the rest of the signals. In the case of long distances between Alice and Bob, the losses in the communication channel become so high that blocking part of the signals in these attacks causes no extra losses, and protocols can not provide secret key any more. Practical issues in decoy-state quantum key distribution were studied~\cite{FedorovPRA2017}.

There exist various QKD protocols that promise higher key rate over longer distances and robustness against attacks like PNS and USD. One of these protocols is the differential phase shift (DPS) protocol~\cite{DPS_original}, which is in addition relatively easy for practical implementation. In this protocol the relative phase between the adjacent pulses is used to encode the bit value of the key. Specially, DPS uses the train of coherent pulses and information is encoded in the phase difference between any two subsequent pulses. The receiver uses the delayed arm Mach-Zehnder interferometer to superpose adjacent pulses and in dependence on their phase difference he obtains bit value either zero or one. The described encoding does not allow Eve to block one of the pulses in the case of failed measurement, since such blocking causes high error rate at Bob side. Thus this protocol is robust to attacks on single transmitted states.

There are several versions of the DPS protocol. Here we consider the one for which the security has been proven~\cite{DPSs1,DPSs2}. We propose a new attack on this version of the protocol whose idea is based on soft filtering operation~\cite{KM_JETP,K_LP} together with proposed earlier active beam splitting attack~\cite{ABS}. Eve takes a part of the states, performs information extraction at every position in the train and then makes a decision, whether to send the rest of the states to Bob or to block it. Such a scenario allows Eve to make a decision, but it does not spoil the states that reach Bob in the case if some of Eve's operations are failed. The unambiguous state discrimination operation can be considered as a special case of soft filtering. Therefore the proposed attack should be at least as good as the USD attack.

We use two assumptions: (a) global phase is known to Eve, and (b) Eve can substitute the original channel between Alice and Bob by a lossless channel. These are the common assumptions for attacks on the DPS protocol~\cite{DPS_attack_sequential,DPS_attack_BGS}). Let us note that security proofs given in~\cite{DPSs1,DPSs2} assume that the global phase is not known to Eve. The motivation for this research is to provide a new upper bound for the DPS security under the assumptions listed above, and to design a universal attack scheme which can be effective for both short and long communication distances, since soft filtering operation has in general a higher success probability than unambiguous state discrimination.

Using the above two assumptions, we derive the expression for Eve's information. This expression is then subject to optimization over the parameters which characterize Eve's attack. The criteria for the parameters realizing maximum of Eve's information are found by using analytical and numerical methods. Optimization of properties of quantum systems is related to general field of quantum control~\cite{QCBooks}, that 
can have applications in quantum information technologies,  creation of quantum correlations, evolution of a particle on a torus, using non-classical light, etc.~\cite{RabitsReview2012,VolovichEPL2016,VolovichProcMIAN2016,Glaser2015Report,AntonJMP2017,VolovichPLA2016}. 

This paper is organized as follows. In Sec.~1 we briefly describe the protocol and principles of its operation. In Sec.~2 the details of the proposed attack are presented. Also, we derive the expressions for the information obtained by Eve about the secret key. Finally, in Sec.~3 the quantum bit error rate and the secret key rate are calculated.

\section{Protocol description and notations}
\begin{figure}
\label{DPS_protocol}
\includegraphics[width=1\textwidth]{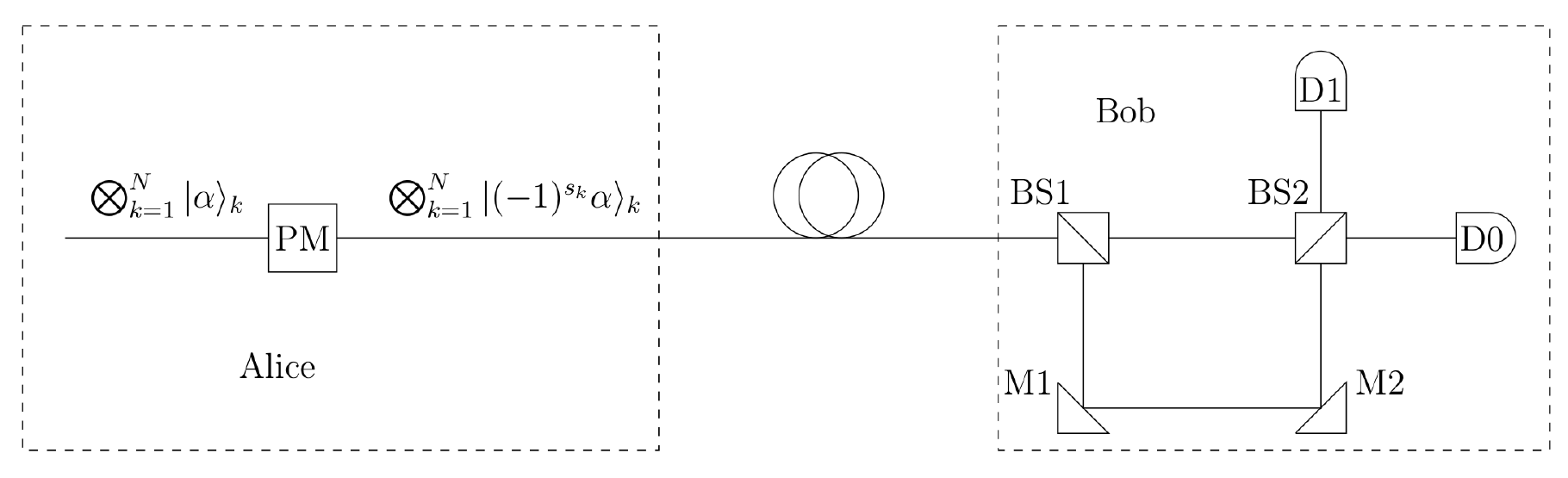}
	
	\caption{Schematic setup of the DPS protocol. Alice produces a train of $N$ coherent states. Then, by using the phase modulator (PM) she transforms her states to the row string $s$. Bob uses Mach-Zehnder interferometer to obtain its bit values. Here BS1 and BS2 denote beam-splitters of the interferometer and M1 and M2 denote its mirrors. Signals of the detectors D0 and D1 correspond to bit values 0 and 1, accordingly.}
	
\end{figure}

Let us give a brief description of the DPS protocol. Its schematics is shown on Fig.~\ref{DPS_protocol}. For details, see~\cite{DPSs2}.

Alice emits trains of coherent pulses of length $N$, where $N \geq 3$ and is typically about 10. Bob uses a delay interferometer with two 50:50 beam splitters and with delay that is equal to the interval between two subsequent pulses. After the interferometer Bob uses two detectors which detect bit values 0 and 1, respectively. The train of $N$ pulses encodes $N - 1$ logical bits.  Alice uses random phase $\varphi$ for the first pulse of the train, and then her phase modulator randomly sets the relative phase of each pulse either to $0$ or $\pi$, where phase $0$ corresponds to $0$ bit value and phase $\pi$ corresponds to $1$. In the sequel, we will set global phase $\varphi$ to zero and we will denote the signal states as $|\pm \alpha\rangle$, where $\alpha > 0$ and $\mu^A = \alpha^2$ is the intensity of the Alice's state $|\alpha\rangle$. As an example, the train of five coherent states
$$
|\alpha\rangle|\alpha\rangle|-\alpha\rangle|\alpha\rangle|\alpha\rangle
$$
corresponds to sending the bit string $0110$.

With such a distributed encoding, the common individual attacks are not effective for the eavesdropper. If Eve obtains an error discriminating states $|\alpha\rangle$ and $|-\alpha\rangle$ and sends the incorrect state to Bob, he is likely to get two errors in the two subsequent pulses. Hence the bit error rate increases. If Eve tries to perform unambiguous state discrimination on every state, she still cannot block the state at position with inconclusive result, because vacuum state can also produce errors at two positions after the interferometer.

Alice and Bob use an optical fiber which in the absence of Eve transforms the coherent input state of intensity $\mu^A = \alpha^2$ to the coherent state with the same phase and a lower intensity $\mu^B$,
\begin{equation}
\label{expected_mu_Bob}
\mu^B= \mu^A\cdot 10^{-\delta\cdot l/10} .
\end{equation}
Here $l$ is the channel's length in kilometers and $\delta$ is the attenuation parameter. In the numerical analysis below we will consider a typical experimental value $\delta = 0.2$ dB/km.

We consider the protocol version~\cite{DPSs2} where Alice and Bob do expect the states of low intensity and thus low probability of detector click. They consider as conclusive only messages with only one detector click in one of $N - 1$ central positions (with vacuum detected on the edges). Otherwise they do block the entire train of pulses. It can be considered as a countermeasure against the attack when Eve sends the states of high intensity if she obtains a lot of information from the block, giving a lot of detector clicks in the positions convenient for her.

The protocol is described by the following steps.

\begin{itemize}
	\item For $M$ times Alice and Bob repeat the following actions:
	\begin{itemize}
		\item Alice generates a random string $S = s_1,\dots,s_N$, where $s_k\in\{0,1\}$, $k=1,\dots, N$.
		\item Alice creates a train of $N$ sequential weak coherent states
		$$
|S\rangle = \bigotimes_{k=1}^N|(-1)^{s_k}\alpha\rangle_k
$$
		and sends it to Bob.
		\item Bob performs interference measurement of the incoming train of coherent pulses. With probability $1-e^{-\mu^B}$ he should obtain a conclusive outcome. Bob consider only events where he detects vacuum in all but one time slots. He notes the detected bit value and announces the time slot $j$ where he obtains unambiguous outcome over an authenticated public channel.
		\item Alice generates her raw key bit as $s_{j+1}\oplus s_j$, where $j=1,\dots, N-1$.
	\end{itemize}
	\item Bob checks the number of conclusive outcomes. If its rate is less than $1-e^{-\mu^B}$, then the legitimate users abort the protocol.
	\item Alice and Bob correct errors in sifted key.
	\item Alice and Bob perform privacy amplification.
\end{itemize}

\section{Active beam splitting attack}

\begin{figure}
\label{attack_setup}
\includegraphics[width=1\textwidth]{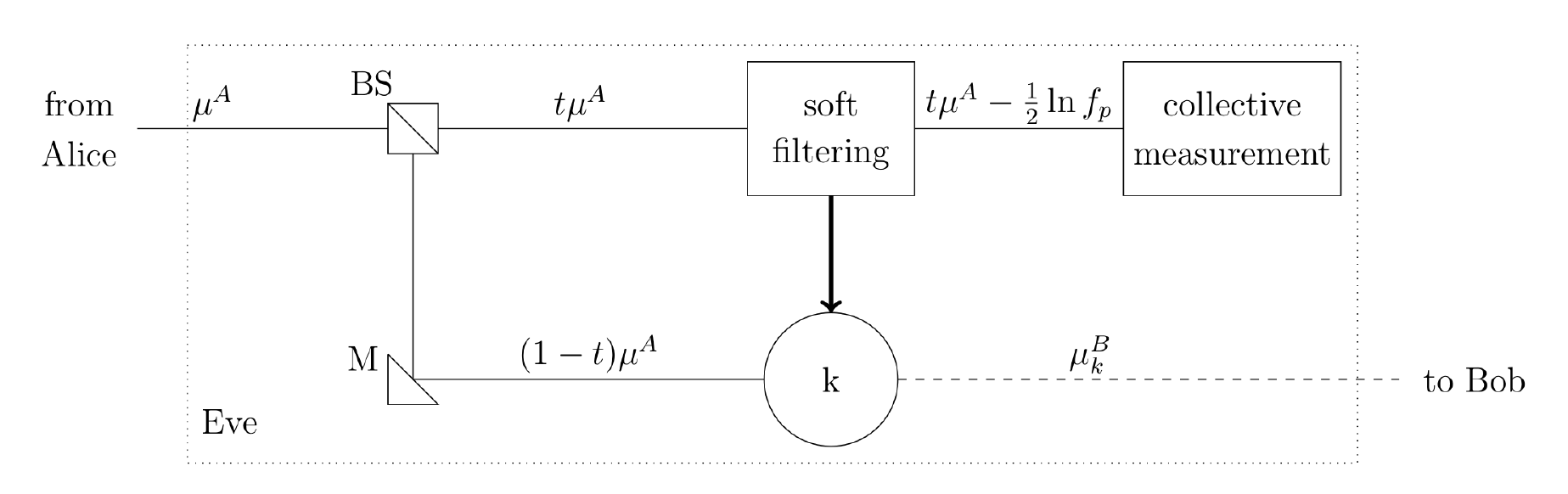}
\caption{Schematic picture of active beam-splitter attack. The Alice's signal of intensity $\mu^A$ comes to the beam-splitter (BS). Eve performs soft filtering operation (SF) on a part of the signal transmitted through BS. Depending on the number of successes $k$ of SF she either sends to Bob states of intensity $\mu_k^B$ or blocks the entire train of states. At the end, she performs a collective measurement to extract maximum information from her states.}	
\end{figure}

Let us design an attack where Eve performs her action on the entire train of coherent pulses. She will sometimes block the entire train and sometimes send a part of original states to Bob via a noiseless channel. Eve should make sure that Bob obtains conclusive results at the same rate as he expects after a lossy fiber channel of the given length.

The scheme of the attack is presented on Fig.~\ref{attack_setup}. It is assumed that Eve is located close to Alice and can insert a lossless channel between her and Bob to send him states without any losses not caused by her beam splitter and blocking decision.

Performing this attack, Eve does divide each state emitted by Alice on her beam splitter into two parts. One has intensity $t\mu^A$ and is used by Eve for information extraction. The other has intensity $(1 - t)\mu^A$ and is used for sending to Bob if Eve decides not to block the states. The decision about either blocking or sending the entire train of pulses to Bob is taken after the  extraction of information by Eve --- this property explains the name of this attack.

The operation that Eve performs on her states $|\pm\widetilde{\alpha}\rangle = |\pm\sqrt t \alpha\rangle$ to extract information is called soft filtering~\cite{KM_JETP} and has the following Stinespring representation
\begin{equation}
\label{soft_filtering}
\begin{split}
|\widetilde{\alpha}\rangle \rightarrow |\psi_0\rangle =  \sqrt{p_s}|\beta\rangle |s\rangle + \sqrt{1 - p_s}|0\rangle|f\rangle, \\
|-\widetilde{\alpha}\rangle \rightarrow |\psi_1\rangle =  \sqrt{p_s}|-\beta\rangle |s\rangle + \sqrt{1 - p_s}|0\rangle|f\rangle,
\end{split}
\end{equation}
where state $|s\rangle$ indicates the successful filtering, state $|f\rangle$ indicates failed filtering, $\langle s|f\rangle = 0$, and $p_s$ is the success probability. By the unitarity condition the success probability equals to
\begin{equation}
p_s = \frac{1 - \langle\widetilde{\alpha}|-\widetilde{\alpha}\rangle}{1 - \langle\beta|-\beta\rangle} = \frac{1 - e^{-2t\mu^A}}{1 - f_p e^{-2t\mu^A}},
\end{equation}
where the filtering parameter $f_p = e^{2(t\mu^A - |\beta|^2)} \in [0, 1]$ is introduced. The filtering parameter equals to zero for unambiguous discrimination of the input states (orthogonal output states $|\pm\beta\rangle$ of infinite intensity; success probability is then $1 - \langle\widetilde{\alpha}|-\widetilde{\alpha}\rangle$), and it equals to one for ``doing nothing'', when output states are the same as input states and success probability is $1$. For the values between $0$ and $1$, this operation increases the distinguishability of the input states (since $\langle\beta|-\beta\rangle < \langle\widetilde{\alpha}|-\widetilde{\alpha}\rangle$) with success probability higher than the one for unambiguous state discrimination. Eve knows whether this transformation was successful or not measuring state of her ancilla, $|s\rangle$ for the successful filtering, and $|f\rangle$ for failed filtering.

After successful soft filtering an optimal collective measurement can be performed to extract maximum information from the output states.

Active beam splitting attack consists of the following steps:
\begin{enumerate}
  \item Eve performs soft filtering operation with filtering parameter $f_p$ on each of her states of intensity $t\mu^A$. This transformation yields a successful outcome with probability $p_s$. In this case Eve gets a lot of information from the amplified states. Otherwise with probability $1 - p_s$ soft filtering fails and Eve gets no information.
  \item 
  If filtering is successful in all positions, Eve in each position has new high intensity states, and can prepare a new high intensity train of coherent pulses for Bob. But there is not always need in it. If the number of successful filtering operations was sufficiently large, Eve sends to Bob states of intensity not higher than $(1 - t)\mu^A$ (which remained after the beam splitter). Starting with some number of failed filtering results in the block, Eve blocks the entire train, since she can extract too little information from it.
\end{enumerate}

Eve's decision on whether to block or not to block a train with a given number of failed filtering operations is based on the requirement that the detection rate of Bob's detectors must be the same as expected without Eve.

If $N$ states of intensity $\mu^B$ do reach Bob without any noise, after the interferometer they are transformed to four states of intensity $\mu^B/4$ on the edges, and $2(N - 1)$ states in the center, which intensity at each position is whether $\mu^B$ or $0$. The protocol requires one detector click in one of the $N - 1$ central positions which has the probability 
$$
(N - 1)e^{-(N - 2)\mu^B}(1 - e^{-\mu^B})
$$
and no clicks for the four states on the edges, which probability is $(e^{-\mu^B/4})^4$. Thus, the probability that Bob has conclusive outcome in this case is given by
\begin{equation}
\label{Bob_conclusive_probability}
p_{\mu^B, N} = (N - 1) e^{-(N - 1)\mu^B}(1 - e^{-\mu^B}).
\end{equation}

If Eve performs her attack, the value of intensity $\mu_k^B$ of signal sending to Bob from Eve in general depends on the number of successful filtering outcomes $k$, but is the same for every state within the train of pulses. Thus Bob can get the train of pulses of intensity $\mu_k^B$ with corresponding probability $p_k$. The probability of blocking the entire train is then $1 - \sum_{k\geq  K} p_k$, where $K$ is the threshold number of successful filtering outcomes: if $k<K$, Eve blocks the entire train. Then, conclusive outcome probability at Bob side in given by
\begin{equation}
\label{Bob_conclusive_probability_total}
p_{{\rm conc}, N}^B = \sum_{k\geq K} p_k p_{\mu_k^B, N}.
\end{equation}
Eve must then choose the values of $K$ and $\mu_k^B$ so that (\ref{Bob_conclusive_probability_total}) is the same as for the expected intensity (\ref{expected_mu_Bob}), i.e. it must be equal to (see (\ref{Bob_conclusive_probability}))
$
p_{\mu^B, N}. 
$

Let us give a more detailed description of Eve's actions.

If Eve receives no failed filtering outcomes, then a train of high intensity (higher, than the original intensity $(1 - t)\mu^A$) can be sent. If the total share of such trains allows Eve to compensate the loss of intensity from blocking other trains, then Eve leaves only them, and the attack gives Eve a lot of information But this is possible only for a large attenuation in the communication channel, that is, for channels of long length. If the length of the channel is such that using only high intensity trains is not enough to compensate the losses, then Eve is forced to send other trains as well, where she did not obtain all the successful filtering outcomes. Then Eve also sends the trains where she obtained one failed filtering result.

If the probability of Bob's conclusive result, taking into account the sent trains of high intensity coherent pulses and trains with one Eve's failed filtering, is exactly the same as expected, then Eve leaves only such trains and blocks all trains with a larger number of failed filtering results. However, it can also happen that the probability of Bob's conclusive outcome after getting the trains of these two types will be \emph{higher} than expected. In this case, Eve artificially reduces the intensity of the trains with one failed filtering, since her task is eavesdropping with maximum of her information. Eve does this by taking at one more beam splitter other small part of the state from which she can also extract information.

These actions can be easily generalized: in the general case, Eve blocks all trains with the number of successful filtering operations less than $K$, reduces the intensity of trains with exactly $K$ successes to $\mu_k^B \leq (1 - t)\mu^A$, sends trains of the intensity $(1 - t)\mu^A$ with the number of successes from $K + 1$ to $N - 1$, and sends trains of higher intensity $\mu_N^B \geq (1 - t)\mu^A$ where filtering succeeded at every position.

After soft filtering Eve performs optimal measurement. Therefore Eve information per one position after obtaining $k$ successful filtering results out of $N$ can be estimated as
\begin{equation}
\label{Eve_information_one_fail}
I_{k}^{AE} \geq \frac{k\chi({|\beta\rangle, |-\beta\rangle}) - 1}{N - 1} =
\frac{1}{N-1} \left\{k\cdot h_2\left(\frac{1 -f_p e^{-2t\mu^A}}{2}\right) - 1\right\},
\end{equation}
where $h_2(x) = -x \log x - (1-x)\log{(1 - x)}$ is the binary Shannon entropy function and the information that Eve can extract from a single position after successful filtering is given by Holevo value~\cite{Holevo} of states $|\pm\beta\rangle$:
$$
\chi({|\beta\rangle, |-\beta\rangle}) = h_2\left(\frac{1 - \langle\beta|-\beta\rangle}{2}\right) = h_2\left(\frac{1 -f_p e^{-2t\mu^A}}{2}\right).
$$

In (\ref{Eve_information_one_fail}) Eve gets information from the states $|\pm\beta\rangle$ at $k$ positions, and at most one bit of information is lost due to distributed encoding by Alice and Bob.

\section{The parameters of the attack and Eve's information}

Let us describe the set of parameters which do characterize the attack.

The first parameter is the part of the state $t$ taken at the beam splitter. For Bob's conclusive probability~(\ref{Bob_conclusive_probability}) there is the optimal value of $\mu^B$, where this probability reaches the maximum. Let us denote this optimal value as $\tilde{\mu}^B$. When Eve takes the part of the state, she should not send to Bob the part of  intensity higher than $\tilde{\mu}^B$, because in this case Bob's conclusive probability becomes lower. Thus $t$ takes values in $[\max\{0, 1 - \tilde{\mu}^B/\mu^A\}, 1]$. It is easy to see that for short channel lengths, Eve can not take most of the state and hence the parameter $t$ has to be relatively small. For a long channel length and large losses expected by legitimate users, the large values of $t$ close to $1$ are optimal. Also, $t$ can take the value $1$. In this case Eve performs filtering over all the states and then sends to Bob a part of her amplified signals only if it succeeds at every position.

The second Eve's parameter is the filtering parameter $f_p \in [0, 1]$ which corresponds to the amplification of the output state for filtering operation. As mentioned above, $f_p=0$ value corresponds to unambiguous state discrimination, and $f_p=1$ corresponds to the output states remaining the same as the input states.

The third and fourth parameters $p_1$ and $p_2$ characterize the set of $N + 1$ intensities $\{\mu_k^B\}$ $(0 \leq k \leq N)$ of the states that are sent to Bob. The first $K$ of them equal zero. The next intensity $\mu_K^B$ takes value from $0$ to $(1 - t)\mu^A$. The following $N - K - 1$ intensities are all equal $(1 - t)\mu^A$. Finally, the intensity $\mu_N^B$ can generally be above $(1 - t)\mu^A$, but usually it does not make sense to make it very high because it would increase the probability that Bob gets two or more detector clicks, and hence would reduce the conclusive probability at Bob's side.

The parameter $p_1 \in [0, N]$ characterizes the threshold number $K$ of
successful filtering results and the intensity of $\mu_K^B$. Integer value of $p_1$ from $0$ to $N - 1$ means that $K = p_1$ and $\mu_K^B = (1 - t)\mu^A$, i.e. Eve blocks all the messages with less than $p_1$ successful filterings and does not change the intensity of the message with $p_1$ successes. For real value $p_1 \in [K, K+1]$ Eve still blocks all the messages with less than $K$ successful filterings and sets $\mu_K^B = (K + 1 - p_1)(1 - t)\mu^A$.

The parameter $p_2 \in [0, 1]$ corresponds to the amplification of the intensity $\mu_N^B$. This intensity should be between the original and optimal values. The parameter value $p_2 = 0$ corresponds to the original value $\mu_N^B = (1 - t)\mu^A$, and $p_2 = 1$ corresponds to $\mu_N^B = \min\{\mu^A - \frac12 \ln f_p, \tilde{\mu}^B\}$. Here, $\mu^A - (\ln f_p)/2$ is the total Eve's intensity after successful soft filtering with parameter $f_p$; obviously, $\mu_N^B$ can not be larger. By linearity, we have
$$
\mu_N^B = (1 - t)\mu^A + p_2 \left(\min\left\{\mu^A - \frac12 \ln f_p, \tilde{\mu}^B\right\} - (1 - t)\mu^A\right).
$$

For each set of these three parameters, one can calculate both Eve's information and the length of the channel at which an attack with this set of parameters is possible. Similarly, for each channel length, one can find the optimal set of parameters, which gives maximum information to Eve.

Eve's information can be found for every value $\mu_i^B$ of intensity for Bob's states, and then one can calculate total Eve's information by taking its mean value with respect to the corresponding conclusive probabilities at Bob's side. The first $K$ parameters obviously do not contribute to Eve's information, since Bob does not obtain a conclusive outcome in this case.

The intensity $\mu_K^B$, if it is less than $(1 - t)\mu^A$, gives Eve information from additional part of the states originally intended for Bob. So in the case of successful filtering, Eve's intensity increases and becomes $\mu_{K, 1}^E = -\frac12 \ln(f_p e^{-2t\mu^A}) + (1 - t)\mu^A - \mu_K^B$. In the case of failure it equals $\mu_{K, 2}^E = (1 - t)\mu^A - \mu_K^B$. Eve's information is then
$$
I_K^{AE} = \frac{1}{N-1}\left\{K h_2\left(\frac{1 - f_p e^{-2(\mu^A - \mu_K^B)}}{2}\right) + (N - K) h_2\left(\frac{1 - e^{-2[(1 - t)\mu^A - \mu_K^B]}}{2}\right) - 1\right\}.
$$
Actually, $I_K^{AE}$ is the lower bound of Eve's information in the case $k=K$.

For the intensities $\mu_k$ between $K + 1$ and $N - 1$ Eve's information lower bound similarly to~(\ref{Eve_information_one_fail}) equals
$$
I_k^{AE} = \frac{k h_2(\frac12(1 - f_p e^{-2t\mu^A})) - 1}{N - 1}.
$$
For the $\mu_N^B$ intensity Eve can give a part of the state to Bob, and her information therefore can decrease. Her intensity after the filtering then equals $\mu_N^E = -\frac12 \ln(f_p e^{-2t\mu^A}) + (1 - t)\mu^A - \mu_N^B$, and her information lower bound is
$$
I_N^{AE} = \frac{N h_2(\frac12(1 - f_p e^{-2(\mu^A - \mu_N^B)})) - 1}{N - 1}.
$$

As a result, for a set of intensities $\{\mu_k^B\}$ with corresponding Eve's successful filtering probabilities
$$
p_k^E = C^k_N p_s^k (1 - p_s)^{N - k} 
$$
taking into account the probability~(\ref{Bob_conclusive_probability}) of obtaining conclusive result by Bob, one has
$$
I^{AE}= \frac{\sum\limits_{k = K}^N p_k^E p_{\mu_k^B, N} I_k^{AE}}{\sum\limits_{k = K}^N p_k^E p_{\mu_k^B, N}} .
$$
This information lower bound can be calculated for each set of parameters, and Eve's goal is to find the optimal parameters for each channel length.

To simplify this computational task, let us derive an analytical relation on the parameters which provide an extrema of the information
\[
I^{AE}(t, f_p,\mu_K^B, \mu_N^B)=\frac{\sum\limits_{k=K}^N Z_k I_k^{AE}}{\sum\limits_{k=K}^N Z_k},
\]
where we denote
\[
Z_k=Z_k(t, f_p, \mu_K^B, \mu_N^B)=C_N^k(N-1) p_s^k(1-p_s)^{N-k}e^{-N\mu_k^B}(e^{\mu_k^B}-1)
\]
and for simplicity consider the intensities $\mu_K^B$ and $\mu_N^B$ instead of corresponding parameters $p_1$ and $p_2$.

Maximum of $I^{AE}$ can be either on the boundary of the set of all admissible values of the parameters $(t, f_p,\mu_K^B, \mu_N^B)$, or inside of this set. In the latter case derivatives of $I^{AE}$ with respect to each variable should be equal to zero. The derivatives over $f_p$ and $t$ give rather complicated equations and we do not write them here. However, setting to zero derivatives over $\mu_K^B$ and $\mu_N^B$ leads to a relatively simple relation for the optimal parameters.

First let us set to zero the derivative with respect to $\mu_K^B$. Note that only $Z_K$ and $I_K^{AE}$ depend on $\mu_K^B$. Therefore
\begin{eqnarray}
\frac{\partial I^{AE}}{\partial\mu_K^B}&=&0\nonumber\\
\Rightarrow 0&=&\left(\partial_{\mu_K^B} Z_K I_K^{AE}+Z_K\partial_{\mu_K^B} I_K\right)\sum\limits_{l=K}^NZ_l-
\partial_{\mu_K^B} Z_K\sum\limits_{l=K}^NZ_l I_l^{AE}\nonumber\\
\Rightarrow 0&=&
\Bigl(\partial_{\mu_K^B} Z_K I_K^{AE}+Z_K\partial_{\mu_K^B} I_K-
\partial_{\mu_K^B} Z_KI^{AE}\Bigr)\sum\limits_{l=K}^NZ_l\nonumber\\
\Rightarrow I^{AE}&=&
I_K^{AE}+Z_K\frac{\partial_{\mu_K^B} I_K^{AE}}{
\partial_{\mu_K^B} Z_K}\label{eq1}
\end{eqnarray}
where we take into account that
\[
\sum\limits_{l=K}^NZ_l >0.
\]

Similarly setting to zero the derivative with respect to $\mu_N^B$ and noting that only $Z_N$ and $I_N^{AE}$ depend on $\mu_N^B$, we get
\begin{eqnarray}
\frac{\partial I^{AE}}{\partial\mu_N^B}=0
\Rightarrow I^{AE}=
I_N^{AE}+Z_N\frac{\partial_{\mu_N^B} I_N^{AE}}{
\partial_{\mu_N^B} Z_N}.\label{eq2}
\end{eqnarray}

Comparing~(\ref{eq1}) and~(\ref{eq2}) we get that optimal parameters should satisfy the equality
\begin{equation}\label{eq3}
I_K^{AE}+Z_K\frac{\partial_{\mu_K^B} I_K^{AE}}{
\partial_{\mu_K^B} Z_K}=I_N^{AE}+Z_N\frac{\partial_{\mu_N^B} I_N^{AE}}{
\partial_{\mu_N^B} Z_N}.
\end{equation}

The derivatives appearing in~(\ref{eq3}) can be easily computed as
\begin{eqnarray}
\partial_{\mu_K^B} I_K^{AE}&=&-\frac{K}{N-1}f_p e^{-2(\mu^A-\mu_K^B)} \log\frac{1+f_p e^{-2(\mu^A-\mu_K^B)}}{1-f_p e^{-2(\mu^A-\mu_K^B)}}\nonumber\\
&&-
\frac{N-K}{N-1}e^{-2((1-t)\mu^A-\mu_K^B)}
\log\frac{1+e^{-2((1-t)\mu^A-\mu_K^B)}}{1-e^{-2((1-t)\mu^A-\mu_K^B)}},\\
\partial_{\mu_K^B} Z_K&=& C_N^K(N-1)p_s^K(1-p_s)^{N-K}e^{-N\mu_K^B} \left(N-(N-1)e^{\mu_K^B}\right),\\
\partial_{\mu_N^B} I_N^{AE}&=&-\frac{N}{N-1}f_p e^{-2(\mu^A-\mu_N^B)} \log\frac{1+f_p e^{-2(\mu^A-\mu_N^B)}}{1-f_p e^{-2(\mu^A-\mu_N^B)}},\\
\partial_{\mu_N^B} Z_N&=& (N-1)p_s^Ne^{-N\mu_N^B} \left(N-(N-1)e^{\mu_N^B}\right).
\end{eqnarray}

One can also consider the critical error up to which the protocol can provide secret key distribution, assuming that Eve is applying the considered attack. This is error rate at which Eve's information becomes equal to Bob's one. It is given by
$$I^{AE} = I^{AB} = 1 - h_2(Q)$$
This error can be caused by equipment failures, or can be introduced by Eve so that she knew more than Bob.

\begin{figure}
\label{pic_for_mu01}
\includegraphics[width=1.1\textwidth]{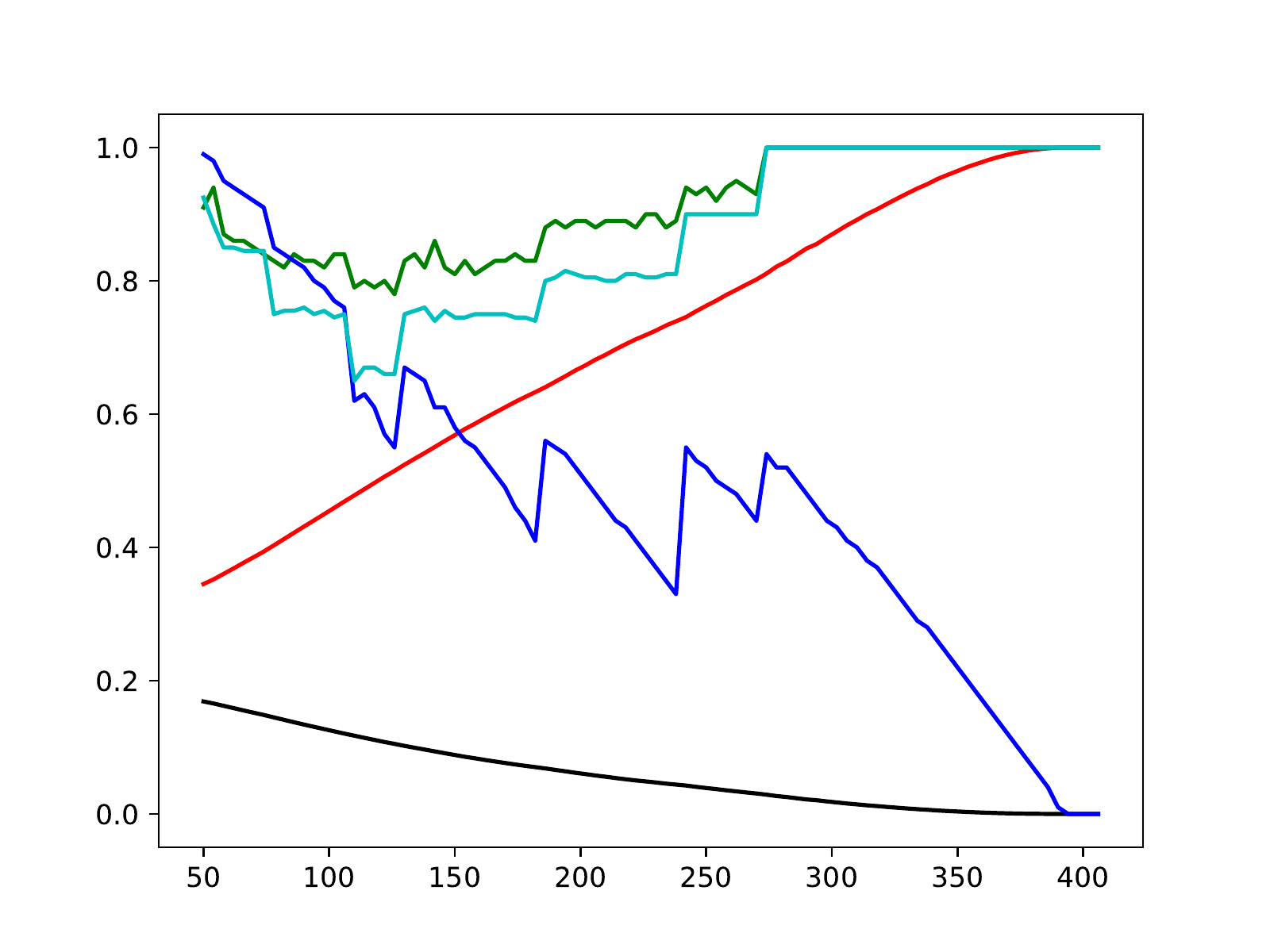}
\caption{Maximum information of Eve (red line) and critical error (black line) for the active beam splitting attack. Block size is $N = 10$, Alice intensity is $\mu^A = 0.1$, and the attenuation parameter is $\delta = 0.2$ dB/km. Horizontal axis shows channel's length (in km). Numerically calculated optimal parameters are also shown: green line for $t$, blue line for $f_p$, and cyan line for $p_1 / N$.}
\end{figure}

Figure~\ref{pic_for_mu01} shows critical error for this protocol for Eve using the considered attack and some of optimal parameter values.

Beside critical error, the secret key rate may be of interest, which depends both on information gap between Bob and Eve and on conclusive result probability on Bob's side. It equals
$$
R = p_{{\rm conc}, N}^B(I^{AB} - I^{AE}),
$$
and it is possible to calculate the optimal value of the initial Alice's intensity $\mu^A$ in order to achieve the maximum secret key rate under the assumption that Eve applies the considered attack.

\begin{figure}
\label{pic_optimal_mu}
\includegraphics[width=1.1\textwidth]{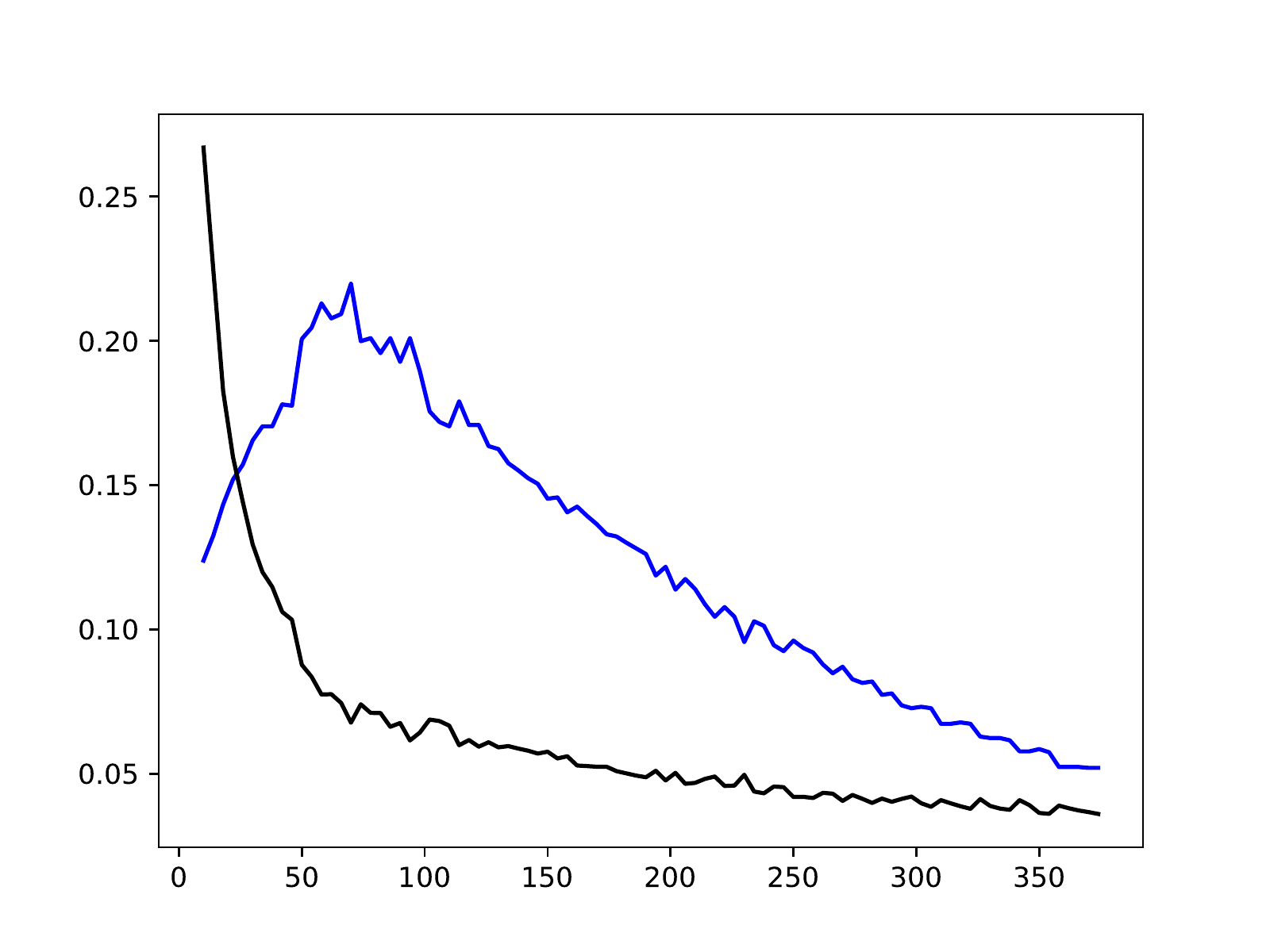}
\caption{Optimal intensity $\mu^A$ of the state produced by Alice (blue line) and critical error (black line) versus channel's length (in km). Block size is $N = 10$ and the attenuation parameter is $\delta = 0.2$ dB/km.}
\end{figure}

Figure~4  shows the optimal value of $\mu^A$ for versus length of the channel (in km).

\section{Conclusions}

We have designed an attack which can be considered as a generalization of both unambiguous state discrimination and beam splitting attacks. As in beam splitting attack, Eve uses a beam splitter to take a part of each state and then to send the rest of the state to Bob. Like in the USD attack, Eve performs information extraction and then makes a decision whether the message should be blocked or not. The operation of information extraction (soft filtering) is a generalization of unambiguous state discrimination: it makes easier to extract information from the states with some non-zero success probability.

The DPS protocol uses trains of pulses, and to have all the information about the sates Eve should have successful results of her information extraction at every position. Nevertheless, the scheme of active beam splitting attack allows Eve to send the unchanged states to Bob in the case when she fails at some positions. This allows Eve to perform this attack for the situations of little attenuation as well.

\section*{Acknowledgements} Development of the attack in Sections 1-3, calculation of Eve’s information and numerical calculation of the optimal parameters in Section 4 were supported by the Russian Science Foundation under grant 17-11-01388 and performed in Steklov Mathematical Institute of Russian Academy of Sciences. Calculation of critical error was performed under grant MK-2815.2017.1. Calculation of the equality for the optimal parameters was performed under the project No. 1.669.2016/1.4.


\begin{thebibliography}{99}

\bibitem{BB84} C. H. Bennett, G. Brassard, Proceedings of IEEE International Conference on Computers, Systems and Signal Processing (1984), 175
\bibitem{Holevo_article} A. S. Holevo, Problems Inform. Transmission, {\bf 9}:3 (1973), 177
\bibitem{Holevo} A. S. Holevo, {\it Quantum Systems, Channels, Information. A Mathematical introduction}, De Gruyter, 2012
\bibitem{Renner_security} R. Renner, {\it Security of quantum key distribution} PhD thesis, quant-ph/0512258 (2005)
\bibitem{FedorovPRA2018} A. S. Trushechkin, P. A. Tregubov, E. O. Kiktenko, Y. V. Kurochkin, and A. K. Fedorov
{\it Phys. Rev. A} {\bf 97} (2018), 012311
\bibitem{PNS_attack} B. Huttner, N. Imoto, N. Gisin, T. Mor, {\it Phys. Rev.
A} {\bf 5}1, (1995) 1863
\bibitem{USD} M. Dusek, M. Jahma, N. Lutkenhaus, {\it Phys. Rev. A} {\bf 62} (1999), 022306
\bibitem{FedorovPRA2017} A. S. Trushechkin, E. O. Kiktenko, and A. K. Fedorov
{\it Phys. Rev. A} {\bf 96}, 022316 (2017).
\bibitem{DPS_original} K. Inoue, E. Waks, Y. Yamamoto {\it Differential Phase Shift Quantum Key Distribution}, {\it Phys. Rev. Lett.} {\bf 89}, (2002), 037902
\bibitem{DPSs1} K. Tamaki, M. Koashi, G. Kato, 2012,  \texttt{arXiv:1208.1995} 
\bibitem{DPSs2} A. Mizutani, T. Sasaki, G. Kato, Y. Takeuchi, K. Tamaki, {\it Quantum Science and Technology}, {\bf 3} (2018), 014003  
\bibitem{KM_JETP} D. A. Kronberg, S. V. Molotkov, {\it JETP} {\bf 118-1}, (2014), 1
\bibitem{K_LP} D. A. Kronberg, {\it Laser Physics}, {\bf 24} (2014), 025202
\bibitem{ABS} D. A. Kronberg, E. O. Kiktenko, A. K. Fedorov, Yu. V,  Kurochkin, {\it Quantum Electronics}, {\bf 47} (2017), 2
\bibitem{DPS_attack_sequential} M. Curty, L. L. Zhang, H-K. Lo, N. Lutkenhaus, {\it Sequential attacks against differential-phase-shift quantum key distribution with weak coherent states}, {\it Quant. Inf. Comput.} {\bf 7} (2007), 665
\bibitem{DPS_attack_BGS} C. Branciard, N. Gisin, V. Scarani, {\it Upper bounds for the security of two distributed-phase reference protocols of quantum cryptography}, {\it New Journal of Physics } {\bf 10} (2008)
\bibitem{QCBooks} D.  D'Alessandro   {\it Introduction to Quantum Control and Dynamics}, 2008, (Boca Raton: Chapman \& Hall); H. M. Wiseman, G.J. Milburn, Quantum Measurement and Control, 2009 (Cambridge Univ. Press:  Cambridge, United Kingdom)
\bibitem{RabitsReview2012}  Brif C., Chakrabarti R. Rabitz H., in Advances in Chemical Physics, edited by S. A.
Rice and A. R. Dinner (Wiley, New York, 2012), vol. 148, p. 1.
\bibitem{VolovichEPL2016} A. S. Trushechkin, I. V. Volovich, Perturbative treatment of inter-site couplings in the local description of open quantum networks, EPL, 113:3 (2016), 30005
\bibitem{VolovichProcMIAN2016} I. V. Volovich, S. V. Kozyrev, Manipulation of states of a degenerate quantum system, \textit{Proc. Steklov Inst. Math.}, {\bf 294} (2016), 241--251
\bibitem{Glaser2015Report} Glaser S J, Boscain U, Calarco T, Koch C P, K\" ockenberger W, Kosloff R, Kuprov I, Luy B, Schirmer S, Schulte-Herbr\" uggen T, Sugny D and Wilhelm F K 2015 \textit{Eur. Phys. J. D} {\bf 69} 279
\bibitem{AntonJMP2017} A.S. Trushechkin, Semiclassical evolution of quantum wave packets on the torus beyond the Ehrenfest time in terms of Husimi distributions, J. Math. Phys., 58:6 (2017), 62102
\bibitem{VolovichPLA2016} I. V. Volovich, Cauchy--Schwarz inequality-based criteria for the non-classicality of sub-Poisson and antibunched light, Phys. Lett. A, 380:1 (2016), 56--58
\end{thebibliography}
\end{document}